\documentclass[11pt]{article}
\usepackage{graphics,amssymb}
\usepackage{epsfig}
 
\begin{document}

\begin{flushright}
{\normalsize
SLAC-AAS-97\\
KEK-ATF-11\\
October 2000}
\end{flushright}

\vspace{.8cm}

\begin{center}
{\bf\Large
Bunch Length Measurements at the\\ ATF Damping Ring in April 2000 
\footnote{Work supported by
Department of Energy contract  DE--AC03--76SF00515.}}

\vspace{1cm}

{\large
K.L.F. Bane,\\ 
\vspace*{3mm}
Stanford Linear Accelerator Center, Stanford University,
Stanford, CA  94309, USA\\
\vspace*{3mm}
T. Naito, T. Okugi, and J. Urakawa\\
\vspace*{3mm}
High Energy Research Organization (KEK),
Oho 1-1, Tsukuba, Ibaraki, Japan

}

\end{center}

\vfill

\title{Bunch Length Measurements at the\\ ATF Damping Ring in April 2000}
\author{K.L.F. Bane, T. Naito, T. Okugi, and J. Urakawa}
\date{}
\maketitle

\section{Introduction}

We want to accurately know the energy spread and bunch length dependence
on current in the ATF damping ring. 
One reason is to know the strength of the impedance:
From the energy spread measurements
we know whether or not we are above the threshold to the microwave instability,
and from the energy spread and bunch length measurements we 
find out the extent of potential-well bunch lengthening (PWBL).
Another reason for these measurements is to help in our understanding of
 the
intra-beam scattering (IBS) effect in the ATF. The ATF as it is now, 
running below design energy and with the wigglers turned off, is
strongly affected by IBS. To check for consistency with IBS theory
of, for example, the measured vertical beam size,
we need to know all dimensions of the beam, including the longitudinal
one.
But beyond this practical reason for studying IBS,
IBS is currently a hot research topic at many accelerators
around the world (see {\it e.g.} Ref.~\cite{CKim}), 
and the effect in actual machines is not well understood.
Typically, when comparing
theory with measurements fudge factors are needed to get agreement
(see {\it e.g.} Ref.~\cite{CKim}).
With its strong IBS effect, the ATF is an ideal machine for studying
IBS, and an indispensable ingredient for this study is a knowledge
of the longitudinal phase space of the beam.

The results of earlier bunch lengthening measurements in the ATF
can be found in Refs.~\cite{Hayano}-\cite{dec99}. Measurements of
current dependent effects, especially bunch length measurements using
a streak camera, can be difficult to perform accurately. For example,
space charge in the camera itself can lead to systematic errors in the
measurement results. 
It is important the results be accurate and
reproducible. In the measurements of both December 1998\cite{dec98} and 
December 1999\cite{dec99}, by using light filters, the authors first checked
that space charge in the streak camera was not significant.
And then the Dec 99 authors show that their results agree
with those Dec 98, {\it i.e.} on the dates of the two measurements
the results were reproducible.

Since IBS is so strong in the ATF, 
in the Dec 99 measurements an attempt was made to estimate the impedance
effect using the following method:
 First, from the form of the energy spread {\it vs.}
current measurements it was concluded that the threshold to the microwave
instability was beyond 2~mA. 
Then, by dividing the bunch length
{\it vs.} current curve by the energy spread {\it vs.} current curve
the effect of IBS was divided out, and 
PWBL was approximated. 
The assumption is that PWBL can be treated as a perturbation
on top of IBS.
The result was that this component
of bunch lengthening was found to grow by 7-15\% (depending
on the rf voltage) between
the currents of .5~mA and 2~mA, about a factor of 3 less than the
total bunch length growth. The conclusion was that the inductive component
of the impedance was small, in fact much smaller than had been concluded
earlier in Ref.~\cite{Hayano}.

Electron machines generally run in a parameter regime where IBS is
an insignificant effect, and impedance measurements and calculations
have also normally been performed for machines where IBS is
 unimportant. 
To simplify the
interpretation of the impedance from bunch length measurements,
in April 2000 
the energy spread and bunch length measurements
of Dec 99 were repeated, but now with 
the beam on a linear (difference) coupling resonance,
where the horizontal and vertical emittances
were approximately equal. For this case the effect of 
IBS was expected to
 be very small. An energy spread {\it vs.} current measurement
under such conditions will also allow us to more clearly see whether
we reach the threshold to the microwave instability. As
part of the April
data taking we,
in addition, repeated the earlier off-coupling measurements,
in order to check the
reproducibility of the earlier results. In this
report we present and 
analyze this recent set of data, and compare it with the 
results of the earlier
measurements, particularly those of Dec 99.  

The measurements and analysis of data in this report follow essentially
the same procedure as was used in Ref.~\cite{dec99}. In the present report
we will try to be relatively brief. The comparison of our results with
IBS theory will be given in a following report.
For more details about the measurement and analysis
techniques presented in this report,
 the reader should consult Ref.~\cite{dec99}.

\section{Energy Spread Measurements}

\begin{figure}[htb]
\centering
\epsfig{file=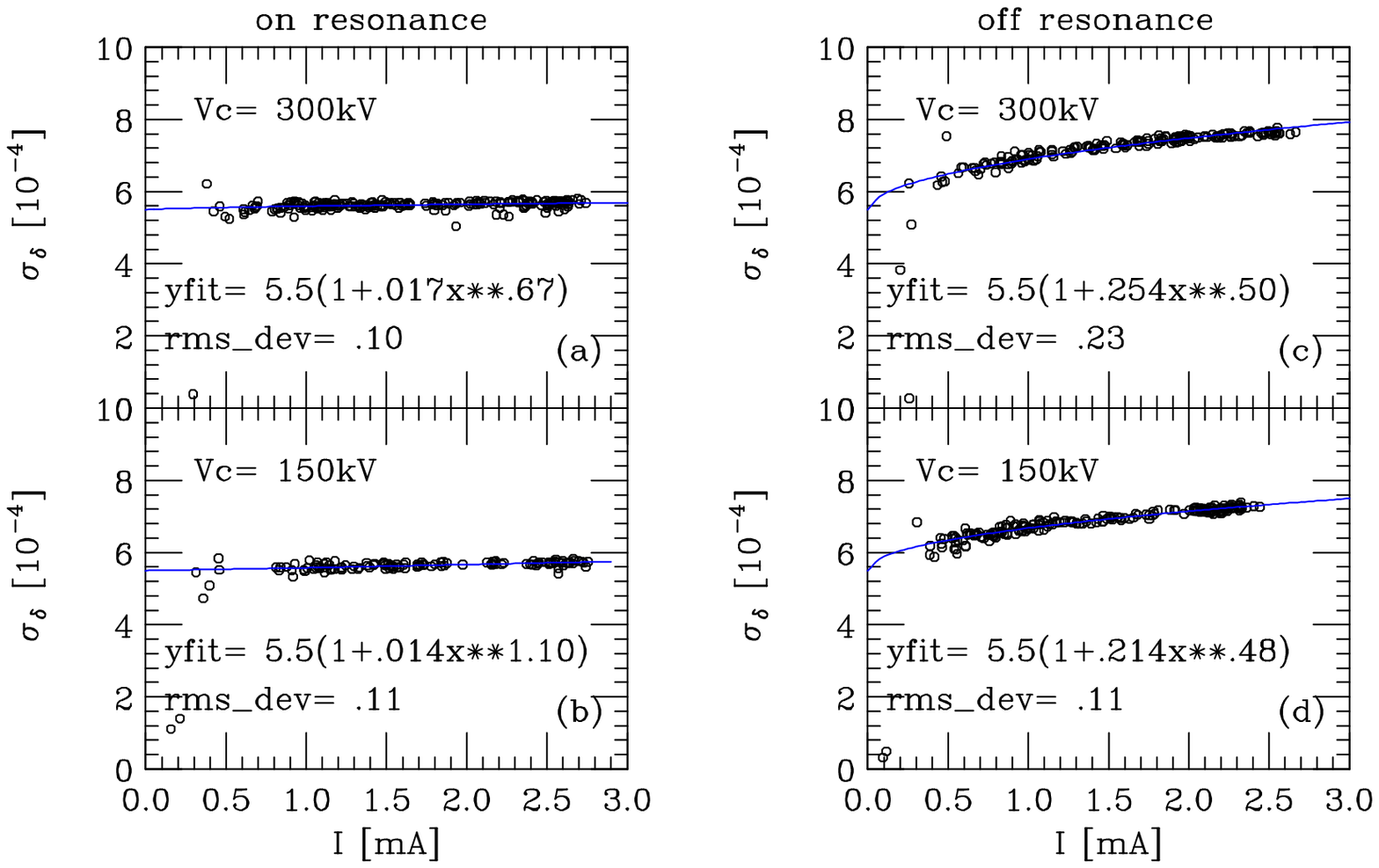, width=12.6cm}
\caption{
Energy spread as a function of current for various values of
cavity voltage.
The measurements were performed on 3/21-22/00. The curves give
a simple fit to the data.
}
\label{fisigdel}
\end{figure}

Energy spread {\it vs.} current measurements were performed
on March 21-22, 2000, with the beam both on and off the coupling
resonance, for rf voltages of $V_c=150$~kV and 300~kV.
As usual the beam width is measured after extraction on a
thin screen in a dispersive region. The data is fit to a Gaussian,
yielding the rms energy spread $\sigma_\delta$ (see
Fig.~\ref{fisigdel}). (For these measurements the energy
spread was not calibrated, but from earlier experience we
expect the zero value results to be near $5.5\times10^{-4}$,
which allows us to estimate the scaling factor.)
In the figure the points in the plots are fit to
$y=5.5(1+a\cdot x^b)$.

We note from Fig.~\ref{fisigdel} that, as expected, on the coupling resonance
the energy spread is almost independent of current, only increasing
by 3-4\% by 2.5~mA. 
With the vertical beam size enlarged, the effect of intra-beam
scattering becomes small.
Off the coupling resonance the change by 2.5~mA
is 35-40\%. 
We also note in (a) and (b)
that there is no evidence of the threshold to the microwave
instability, whose signature would be a kink in the data.
If there is no microwave instability on resonance, then almost certainly
there is no microwave instability off resonance, since in the latter
case the longitudinal phase space volume is increased, which tends
to stabilize the beam. 
Finally, note that the 150~kV and 300~kV on-resonance results are
statistically identical, as would be expected in the case of
no microwave instability and negligible
intra-beam scattering.
Note also that for the off-resonance results the energy spread
is slightly larger for the higher voltage. 
This would be expected, since a higher voltage means a smaller
bunch length, which increases intra-beam scattering, which, in turn,
more increases the energy spread.

\begin{figure}[htb]
\centering
\epsfig{file=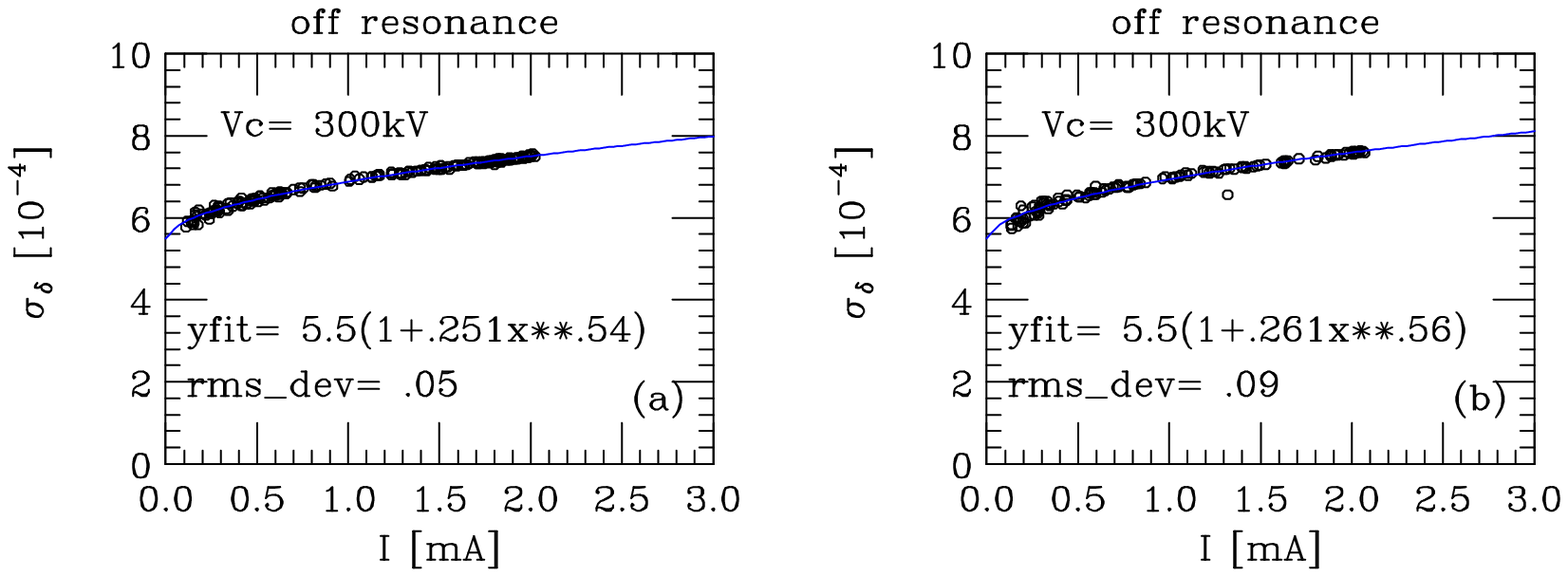, width=12.6cm}
\caption{
Energy spread as a function of current off-resonance, at
$V_c=150$ and 300~kV.
Measurements were performed April 13~(a) and 14~(b).
}
\label{fisigdel_april}
\end{figure}

The measurements were repeated for the nominal (off-resonance, $V_c=300$~kV)
settings on April 13 and 14 (see Fig.~\ref{fisigdel_april}),
 just before and after the bunch length
measurements to be presented in the next section.
We note that the results of the two measurements are essentially the
same, and they are also essentially identical to the March results
with the beam off-resonance and $V_c=300$~kV (Fig.~\ref{fisigdel}c).
This indicates that the machine conditions, as concerns intra-beam
scattering
(the horizontal and vertical beam sizes; the lattice),
are the same during all these measurements, as well
as during the off-coupling, bunch length measurements presented below.

\begin{figure}[htb]
\centering
\epsfig{file=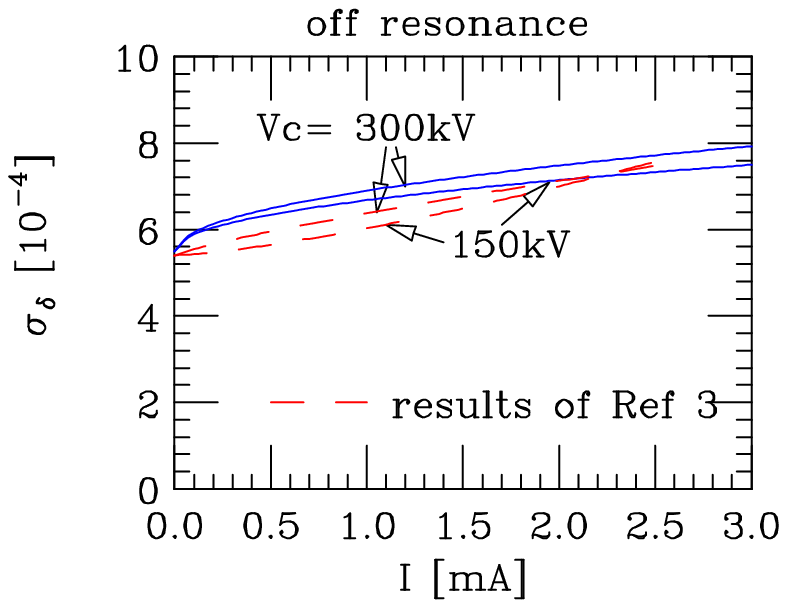, height=4.8cm}
\caption{
Fits to the March, off-resonance energy spread measurements, for
$V_c=150$ and 300~kV, compared to
results of Dec 99 (the dashed curves).
}
\label{fisigdelo}
\end{figure}

Finally, in Fig.~\ref{fisigdelo} we compare the fits to the
$V_c=150$ and 300~kV, off-coupling measurements with 
those to measurements---under the same conditions---taken
in Dec 99 and presented in Ref.~\cite{dec99}.
We note that the corresponding curves for the two dates are
significantly different. For example, for $V_c=300$~kV and
$I=1$~mA the new results are 8\% higher than the earlier results.
This suggests that (when off coupling)
the machine, concerning intra-beam scattering,
was significantly different for the two dates.

\section{Bunch Length Measurements}

For the streak camera measurements
first, to make sure that space charge in the camera itself was not a problem,
tests were made at the highest current (2.5~mA) with different light filters, and an appropriate filter was chosen.
Then, the main data taking process consisted of
storing a high current beam
and measuring the longitudinal bunch
profile $\sim 50-70$ times at fixed time intervals, while
the current naturally decreased. 
Each trace, along with
its DCCT current reading, was automatically saved to disk.
To let in more light at the lower currents
the light filter was automatically
switched half-way through each data taking sequence.
Measurements were
performed with the beam on and off the coupling resonance, and
with 4 different settings of rf voltage.

\begin{figure}[htb]
\centering
\epsfig{file=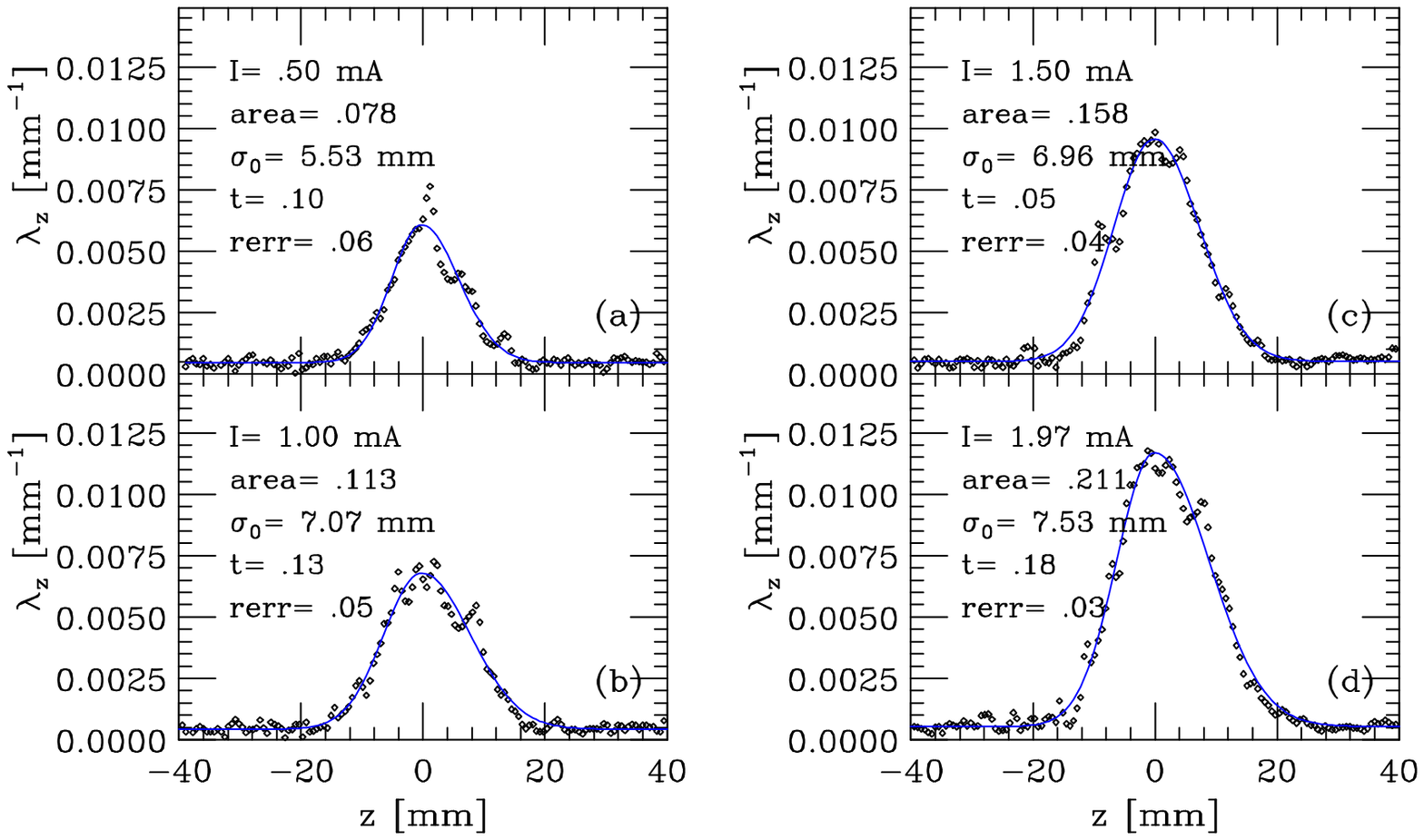, width=12.6cm}
\caption{
Four example scans, and their asymmetric Gaussian fits, for 
the beam on the coupling resonance with $V_c=250$~kV.
The horizontal axis has been shifted so that all peaks are at $z=0$.
}
\label{fiscans}
\end{figure}

As before
the streak camera traces were fit to an asymmetric Gaussian, given by
\begin{equation}
  \lambda_z= {A\over\sqrt{2 \pi}\sigma_0} 
     \exp\left[ -{1\over2} {( z-{\bar z} )^2\over\sigma_0^2(1\pm t)^2}
 \right]\quad\quad\quad z\gtrless{\bar z}\quad,
\end{equation}
with the convention that  
more negative values of $z$ are more to the front of the bunch.
 The fitting constants are $A$, $\sigma_0$, $\bar z$, and
the asymmetry factor $t$ (a constant, platform offset is also included in
the fit). Note that the full-width-at-half-maximum (FWHM) of the fit
is $ z_{fwhm}= 2\sqrt{2 \ln2}\,\sigma_0$
and the rms bunch length
 $ \sigma_z= \sigma_0\sqrt{1 + (3-8/\pi)t^2 }$. For small $t$,
$\sigma_z\approx\sigma_0$.
The skew moment, defined by 
$s=\langle (z-\langle z\rangle)^3\rangle/\sigma_z^3$,
 is given by $s\approx4t/\sqrt{2\pi}$, for small $t$.
Note also that from physical considerations, we generally expect
 $t>0$, {\it i.e.} the leading
edge of the distribution to be steeper than the trailing edge.
Four example scans, with their fits, are shown in Fig.~\ref{fiscans}.
We note that the fits are reasonably good, though we see some, what are
likely anomalous, deviations from the fits in the data.
We have lots of data and we will use the statistical Method of Maximum 
Likelihood to do the error analysis\cite{Bevington}.

\begin{figure}[p]
\centering
\epsfig{file=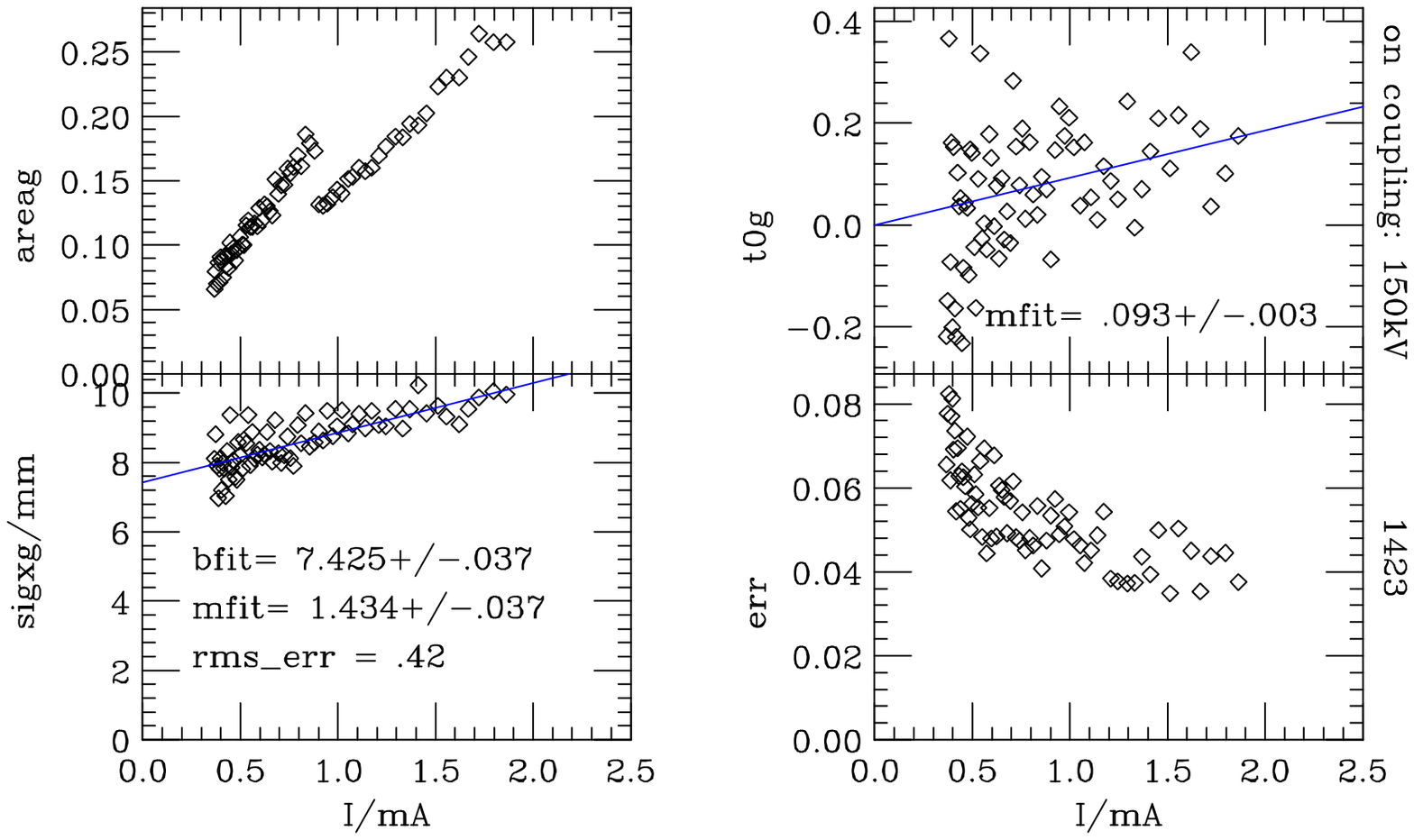, width=12.3cm}
\caption{
Bunch length as function of current for 
the beam on the coupling resonance and $V_c=150$~kV.
Given are the parameters of the asymmetric Gaussian fit
to the measured profiles.
The lines in the plots are straight line fits to the results
(with zero offset in the case of the asymmetry parameter).
}
\label{fisigz1}
\end{figure}

\begin{figure}[p]
\centering
\epsfig{file=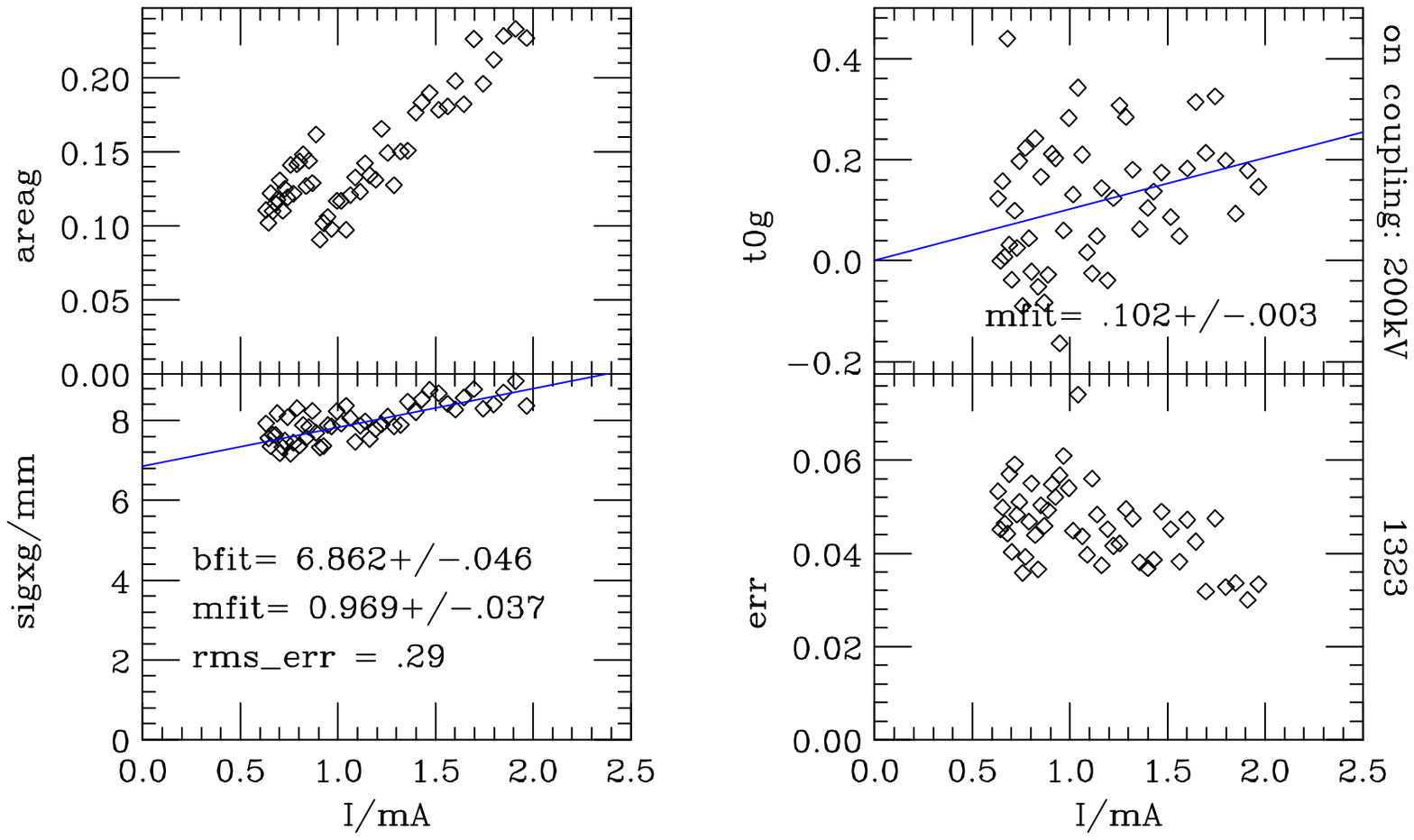, width=12.3cm}
\caption{
Bunch length as function of current for 
the beam on the coupling resonance and $V_c=200$~kV.
}
\label{fisigz2}
\end{figure}

\begin{figure}[p]
\centering
\epsfig{file=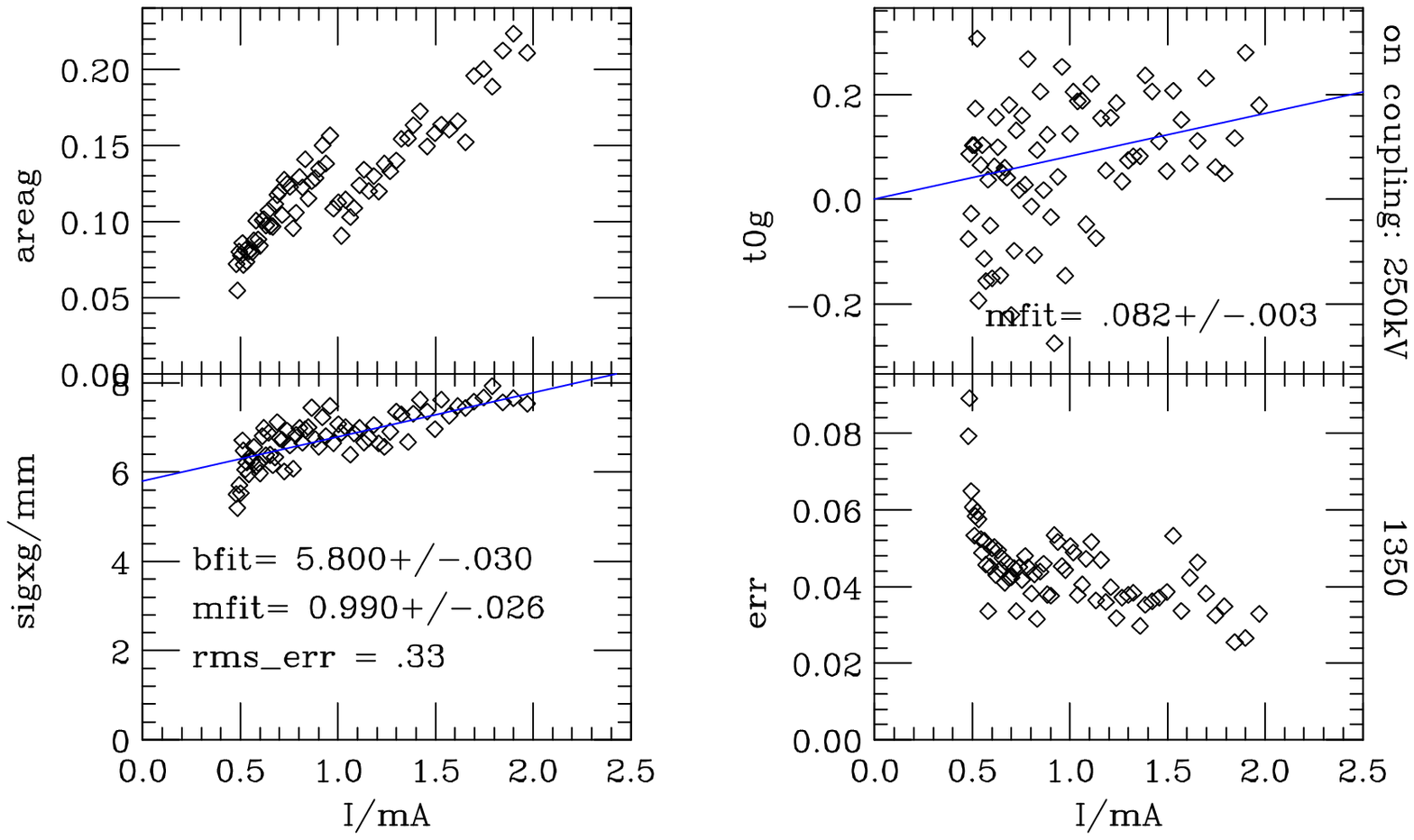, width=12.3cm}
\caption{
Bunch length as function of current for 
the beam on the coupling resonance and $V_c=250$~kV.
}
\label{fisigz3}
\end{figure}

\begin{figure}[p]
\centering
\epsfig{file=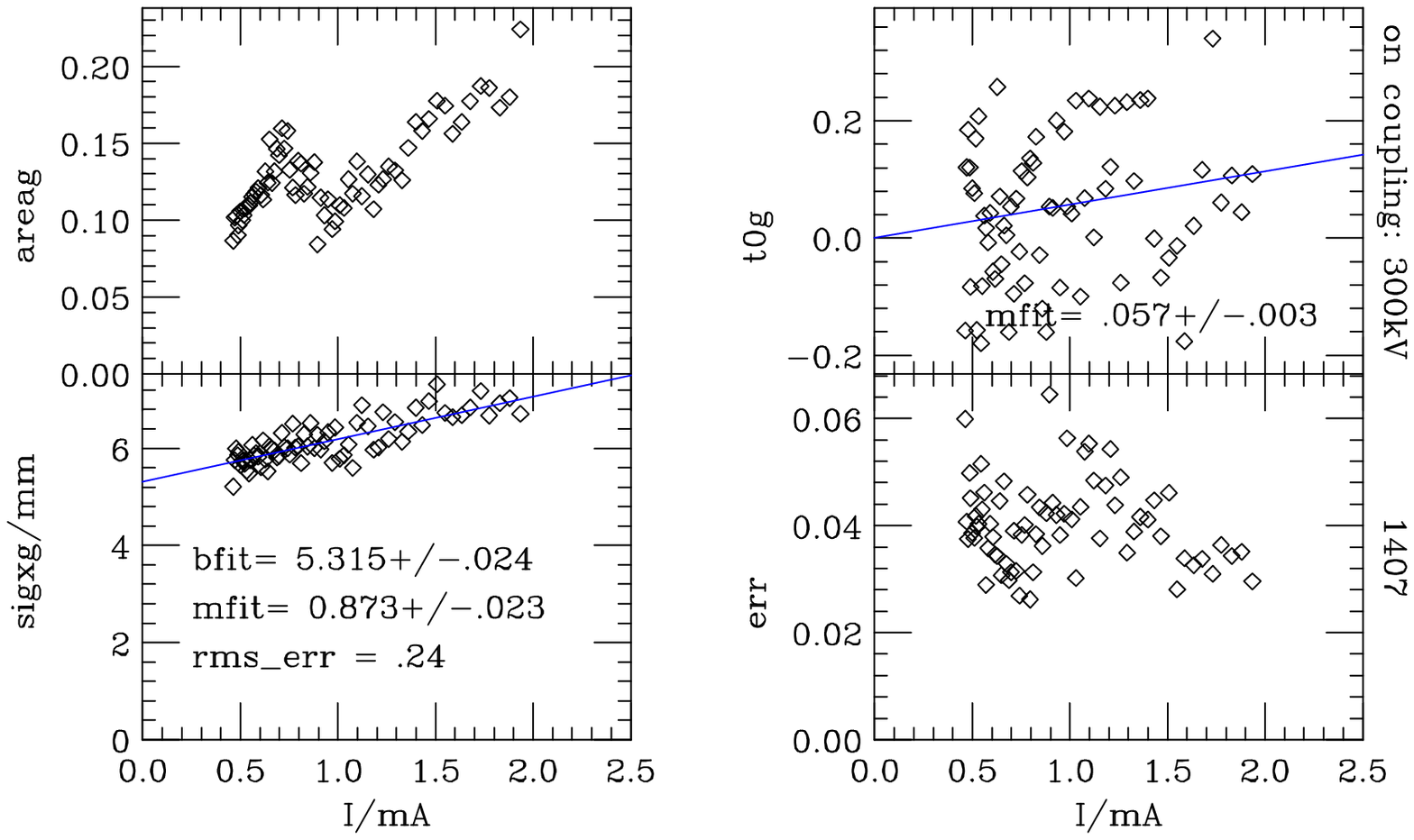, width=12.3cm}
\caption{
Bunch length as function of current for 
the beam on the coupling resonance and $V_c=300$~kV.
}
\label{fisigz4}
\end{figure}

\begin{figure}[p]
\centering
\epsfig{file=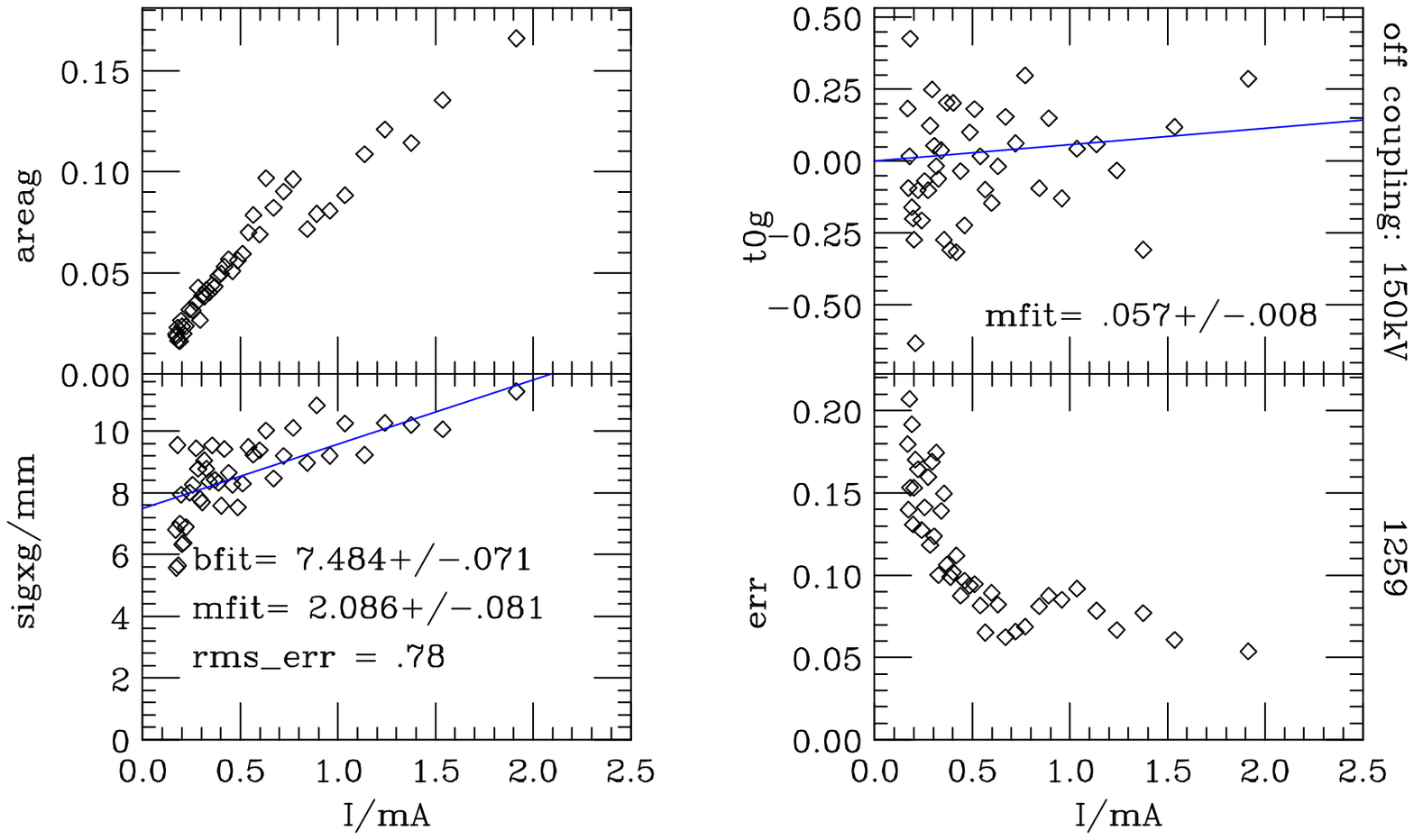, width=12.3cm}
\caption{
Bunch length as function of current for 
the beam off the coupling resonance and $V_c=150$~kV.
}
\label{fisigz1b}
\end{figure}

\begin{figure}[p]
\centering
\epsfig{file=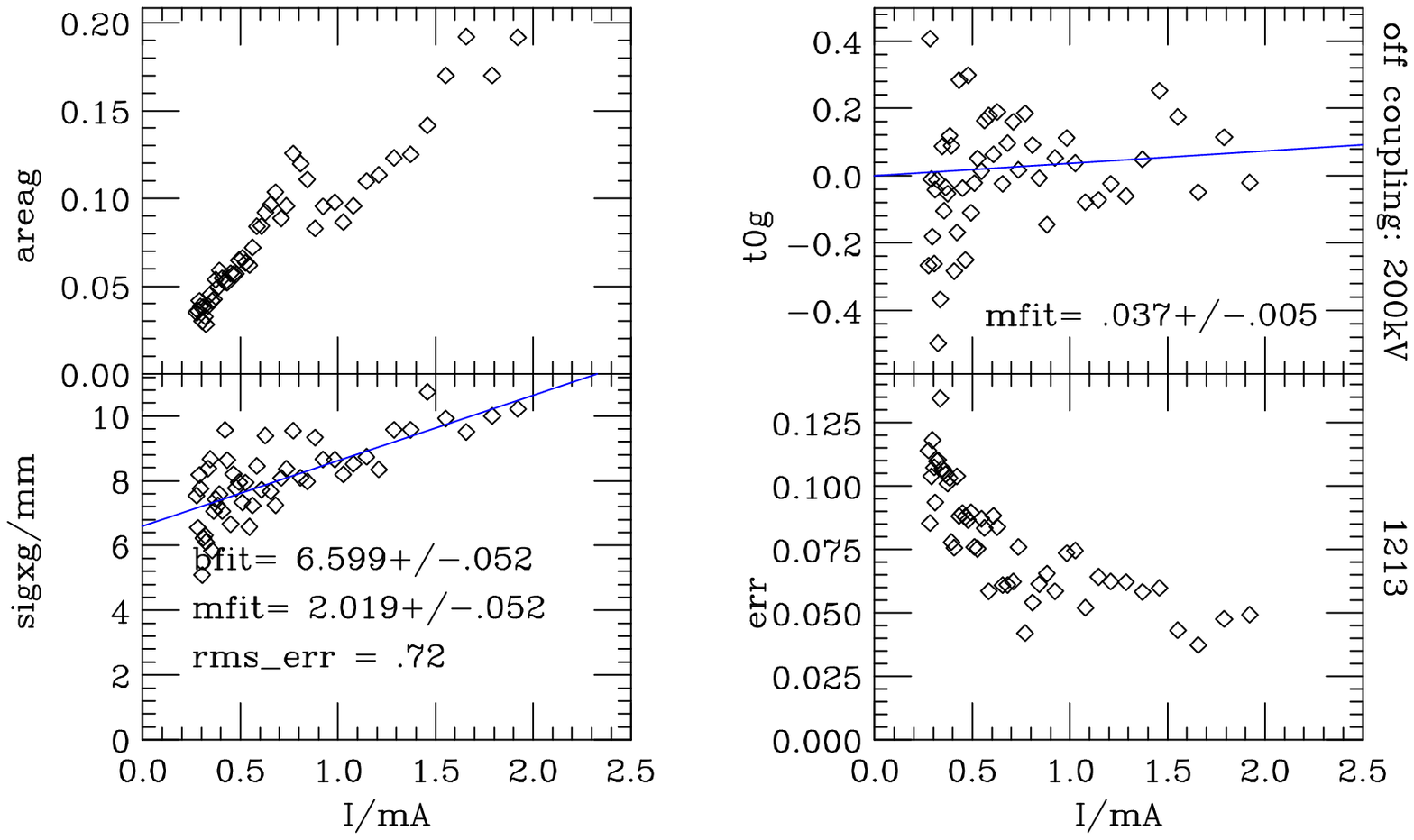, width=12.3cm}
\caption{
Bunch length as function of current for 
the beam off the coupling resonance and $V_c=200$~kV.
}
\label{fisigz2b}
\end{figure}

\begin{figure}[p]
\centering
\epsfig{file=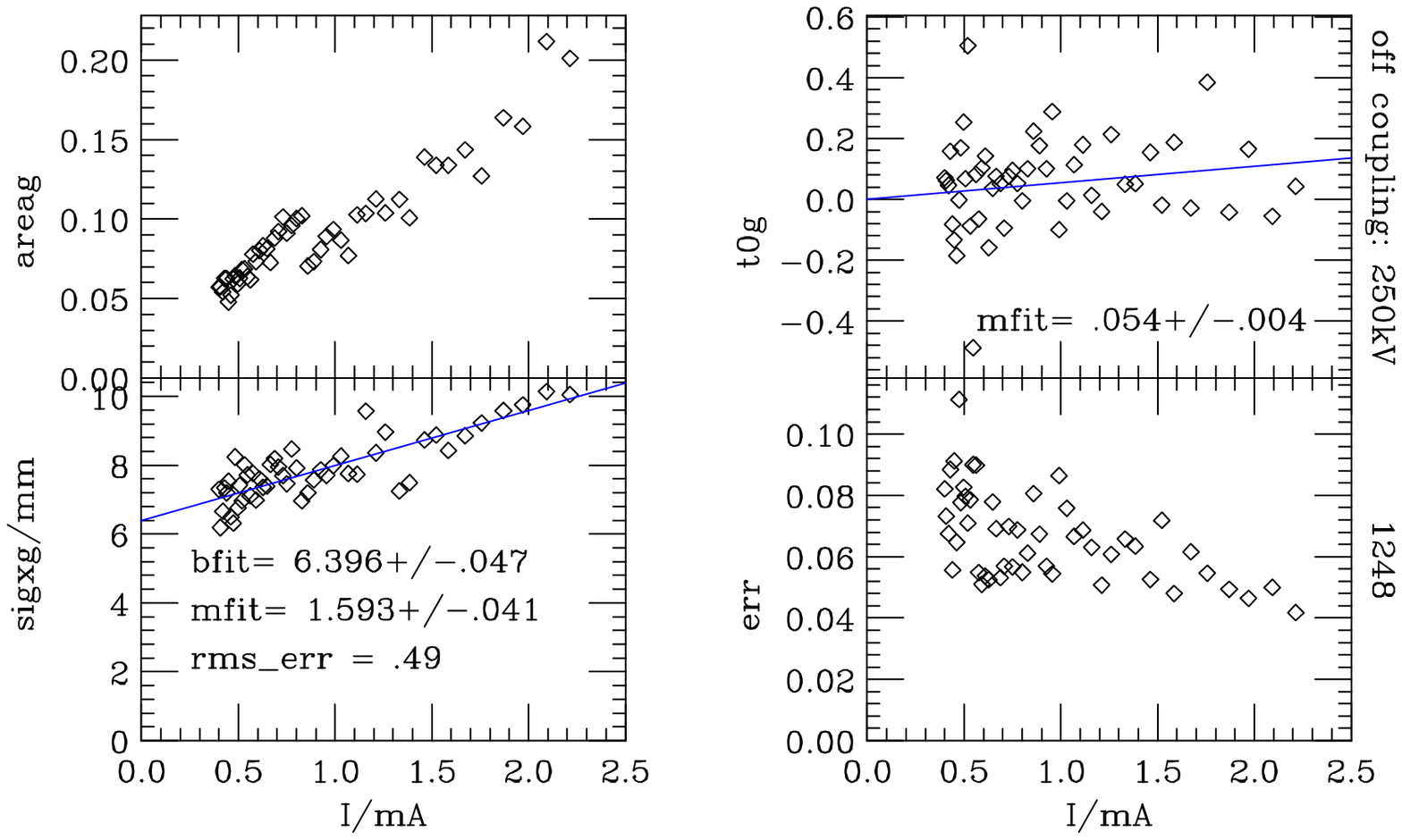, width=12.3cm}
\caption{
Bunch length as function of current for 
the beam off the coupling resonance and $V_c=250$~kV.
}
\label{fisigz3b}
\end{figure}

\begin{figure}[p]
\centering
\epsfig{file=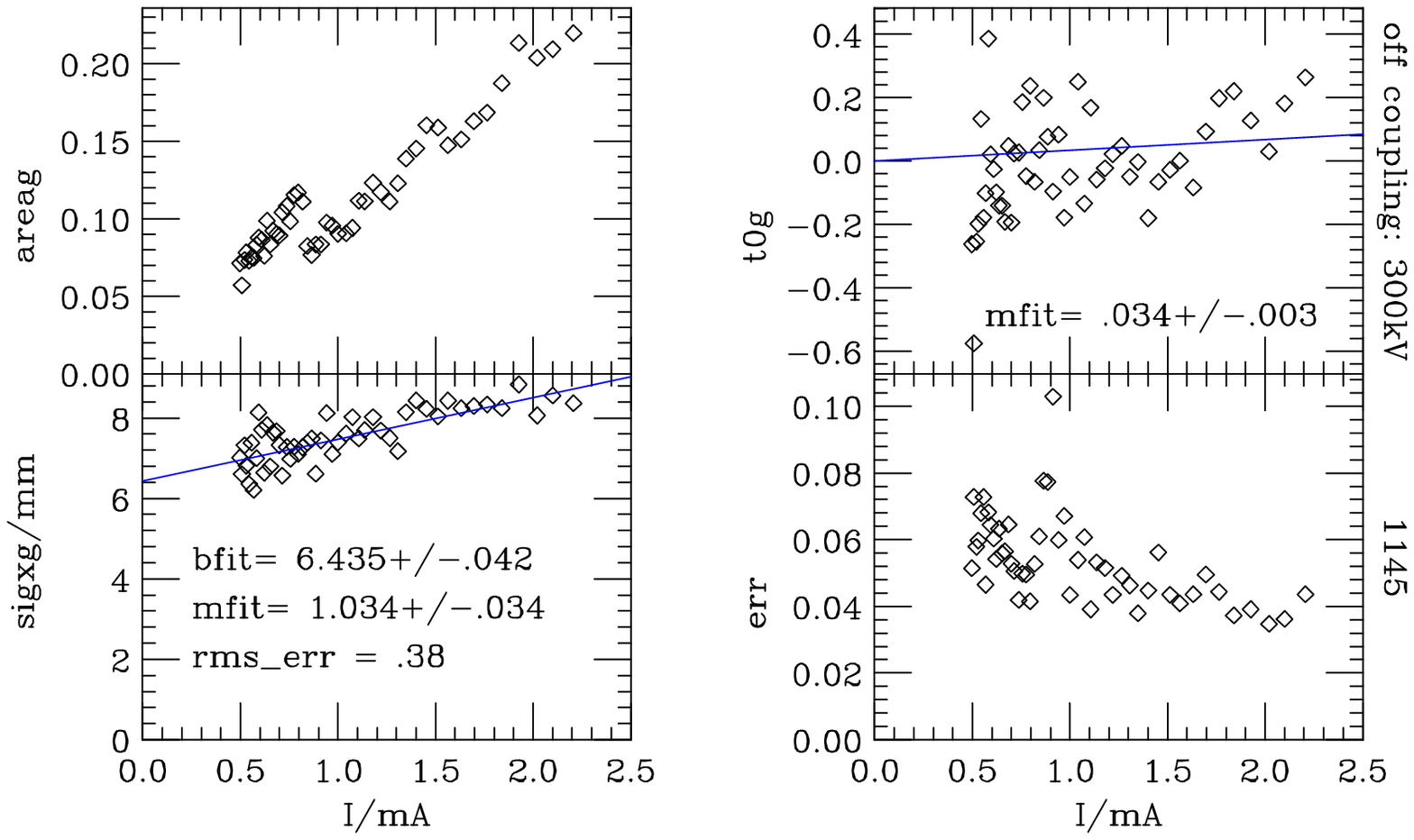, width=12.3cm}
\caption{
Bunch length as function of current for 
the beam off the coupling resonance and $V_c=300$~kV.
}
\label{fisigz4b}
\end{figure}

The results for 
the beam on the coupling resonance and for
rf voltages $V_c= 150$, 200, 250, 300~kV are given
in Figs.~\ref{fisigz1}-\ref{fisigz4}, and for the beam
off the coupling resonance in Figs.~\ref{fisigz1b}-\ref{fisigz4b}.
Shown are the parameters of the asymmetric Gaussian fit
to the measured profiles: the area~(a), the rms $\sigma_0$~(b), 
the asymmetry factor $t$~(c), and an estimate of
the relative rms error in the fit~(d).
Plots (a)-(c) give the zeroth, the second, and the third
moments of the charge distribution.
The line in (b) is a straight line fit to the bunch length, where each
data point has been weighted inversely by the variance
in the fit to the asymmetric Gaussian [note that 
for the fit $y=x\cdot mfit+bfit$].
The line in (c) is a straight line fit to $y=x\cdot mfit$.

In the (a) part of the figures 
we note that the area is roughly proportional to  
current---with 
the constant of proportionality different for the low and the
high current results due to the two different light filters 
used---though with some scatter in the data.
In (b) we note that the scatter in the measured bunch lengths
is quite significant, especially at the
lower currents, with an rms deviation from the straight
line fit varying from .35-.80~mm.
In (c) we see even more scatter in the asymmetry factor.
We expect this parameter to start at zero, and to increase
with increasing current, and the data 
does seem to support an increasingly positive
asymmetry factor with current.
The results are near .1/mA, and slightly decrease with voltage.
We will show in a future ATF report that this parameter
gives us information about 
the higher mode losses and the real part of the impedance.
Note that the measured asymmetry factors off-resonance
are about half of those on-resonance, qualitatively
consistent with the bunch being longer, and the 
higher mode losses being less, when off-resonance.
Finally, in (d) we note that the relative rms error in the
fit to the asymmetric Gaussian 
is $\sim$4-10\%. 
There is more noise than in the measurements of Dec 99,
reflected in the fact 
that the fits then were twice as good.

\subsection{Comparisons and Discussion}

\begin{figure}[htbp]
\centering
\epsfig{file=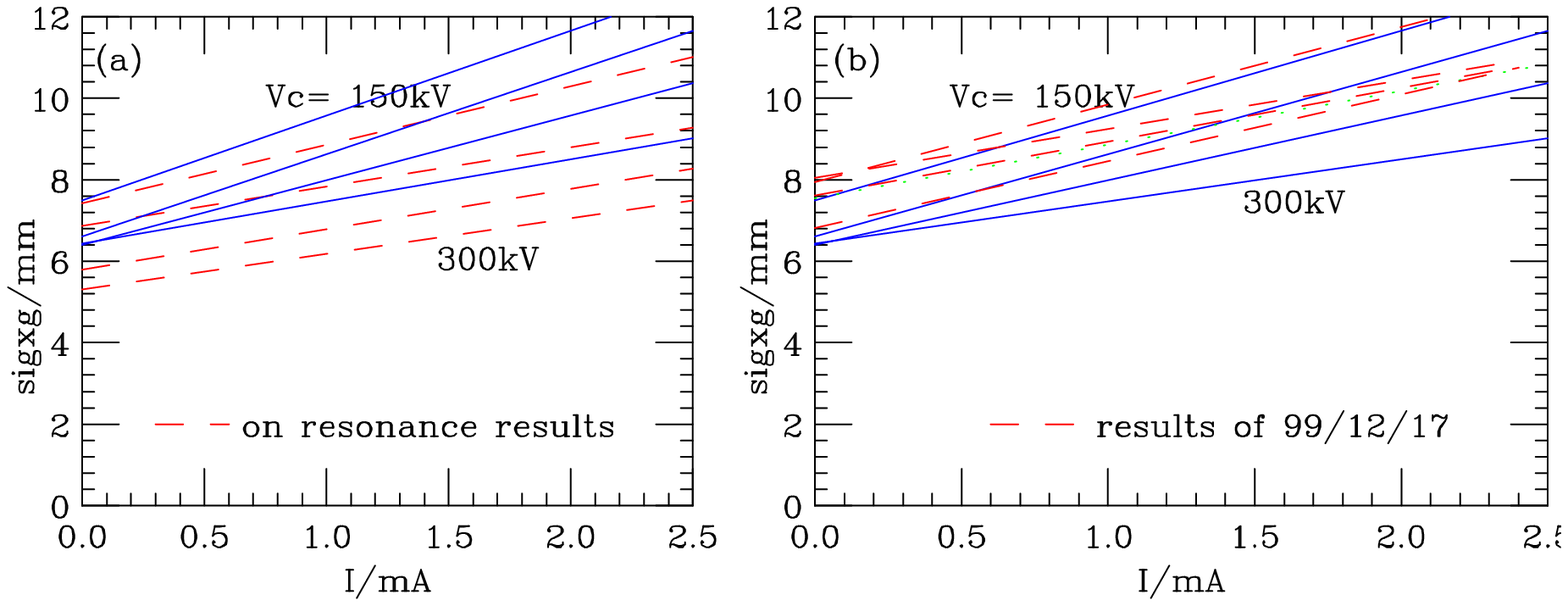, width=12.3cm}
\caption{
Linear fits to measurements with the beam off the coupling
resonance (solid lines). 
The curves represent results for $V_c=150$~kV,
200~kV, 250~kV, and 300~kV.
The dashes in (a) are the on-coupling results; those 
in (b) are results
of Dec 99.
Note that for each case the bunch length monotonically decreases
as the voltage is increased. 
}
\label{fisigzcomp}
\end{figure}

In Fig.~\ref{fisigzcomp} we reproduce the linear fits to the  
bunch length parameter for the measurements
with the beam off the coupling resonance (the solid
lines).
Note that these fits are not valid below $I\lesssim.4$~mA, where there
is no data; in fact, the bunch lengths are expected to drop below
these lines for lower currents.
In (a) the results are compared with the on-coupling results; in
(b) with off-coupling results of Dec 99.
In all cases the bunch length curves move up as the voltage moves down,
as expected. 
Also, in (a) the bunch length is larger off the coupling resonance than
on, due to the effect of intra-beam scattering, as expected.

We expect $\sigma_z(0)$, for $V_c=200$~kV, to be 6.2~mm.
For the case of on-resonance, we find the value at the origin of the
curves to be 3.7\%, 10.4\%, 4.6\%, 5.0\% larger than the zero
 value calculation
for $V_c=150$, 200, 250, 300~kV, respectively. For the case of 
off-resonance the values are 4.5\%, 6.4\%, 15.3\%, 27.1\% larger than
the zero value calculations.
We expect the real bunch lengths to follow curves with negative 
curvature, so an overshoot is not unexpected. 
Note that these overshoots are significantly less than the
$\sim 35\%$ for the Dec 99 results (see Fig.~\ref{fisigzcomp}b),
which, at the time,
 made us suspect a scaling error in the streak camera
results.

In Sec. 2
 we saw that, for the on-coupling case, the energy spread
was almost independent of current. For example, by $I=2$~mA the
energy spread has grown by only 3\%. Here we see that at $I=2$~mA
the bunch length has grown by 33-39\%, depending on the
rf voltage. Since IBS on the coupling resonance
is very weak, this growth must be almost entirely due to PWBL.
Thus, it appears that, after all, PWBL is
a big factor in the ATF, and the machine must have a large inductive
component of impedance.
Note that this conclusion is 
different from that reached in Dec 99, where PWBL
was estimated by dividing the bunch lengthening off-resonance
 by the energy spread growth. There it was found, for
example, that the PWBL contribution to the bunch length at 2~mA,
divided by the contribution at .5~mA, ranged from 7-15\% depending
on voltage. Here the equivalent (on-coupling)
results range from 18-26\%, a marked
difference.
Or, if we try to compare the present on and off-coupling results using
the same prescription, we find that off-coupling the results are
increases of
21.4\% and 6.1\%, for respectively the case $V_c=150$ and 300~kV;
on-coupling the results are 23.5\% and 20.1\%. These also don't agree.
In making the earlier approximate calculation of PWBL we had assumed that
PWBL can be treated as a perturbation on top of IBS. That is, 
we assumed that
(1)~PWBL doesn't affect IBS significantly and (2)~the 
bunch length increase
due to PWBL can be added on top of that due to IBS, and
the result would be similar
whether on or off resonance.
It appears that these assumptions taken together
are wrong and the interaction
of IBS and PWBL is, in fact, more complicated.

In Fig.~\ref{fisigzcomp}b we note that the
 new off-coupling results are very
different from those of Dec 99.
We saw in Sec. 2
above that the energy spread {\it vs.} current was
different for the two measurement days, presumably due to a difference in the
IBS effect, so we would expect differences also in bunch length,
since it is also affected by IBS. In addition the
bunch length (below threshold)
is affected by the impedance through PWBL.
To say more, quantatively, about such measurements 
we would need a better understanding of how IBS and PWBL interact
and, in addition,
 a knowledge of the transverse
beam sizes and optics on the days of the measurements.

\section{Conclusions}

We have performed energy spread and bunch length measurements
with the beam on and off the coupling resonance.
Our energy spread results 
show that, with the beam on-resonance, the effect of
intra-beam scattering is small, and that the
threshold to the micro-wave instability is beyond 2.5~mA.
Our on-resonance bunch length results  show that 
potential well bunch lengthening is large---by 2~mA the
bunch has lengthened by 33-39\%---indicating that the
impedance has a large inductive component.

Our energy spread and bunch length measurements with the beam
off-resonance give results that are very different from those
of Dec 99. This suggests that the horizontal and vertical emittances
and/or the lattice were different than during the earlier measurements,
resulting in a different intra-beam scattering effect.
These measurements should be repeated both on and off the coupling
resonance, under the same conditions---verifiable
by finding the same energy spread {\it vs.} current curves---to
see if the bunch length measurements are reproducible.
We have assumed here that the streak camera measurements are sufficiently
accurate for our purposes. The results appear to be consistent, 
but we have no independent way of checking their accuracy.
(We might suggest installing in the ATF, sometime in the future, a
bunch length measuring apparatus that works in a completely
different way, such as the spectrum approach of Ref.~\cite{Ieiri},
as an independent check on the streak camera results.)

In the future, when doing any such current-dependent measurements
the energy spread {\it vs.} current measurement
should always be performed,
as a specifier of the intra-beam scattering machine conditions,
and since it is a rather quick and simple measurement.
To study specifically impedance questions, measurements should
be performed with the beam on the coupling resonance, where
intra-beam scattering is very weak.
With the beam off-resonance,
bunch length and energy spread measurements are important ingredients,
along with the horizontal and vertical 
emittances and optics, for attempting 
a full understanding of intra-beam scattering
and its interaction with the impedance in the ATF damping ring. 
 
\section{Acknowledgements}

One of the authors (K.B.) thanks the ATF scientists and staff
for their hospitality and help during his visits to the ATF,
and M.~Ross for his encouragement and support to make such visits.


\begin{thebibliography}{99}

\bibitem{CKim}
C. Kim, ``A Three-Dimensional Touschek Scattering Theory,''
LBL-42305, September 1998.

\bibitem{Hayano}
H. Hayano, {\it et al}, ``Impedance Measurement of ATF DR,''
Proc. of EPAC 1998, Stockholm, Sweden, May 1998, p.~481.

\bibitem{dec98}
K. Bane, {\it et al}, ``Bunch Lengthening and Current-Dependent
Energy Spread at ATF,'' ATF Report 98-38, December 1998.

\bibitem{dec99}
K. Bane, T. Naito, T. Okugi, ``Bunch Length Measurements at the
ATF Damping Ring in 1999,'' ATF Report 00-05, May 2000.

\bibitem{Bevington}
See, for example, P. Bevington and K. Robinson,
{\it Data Reduction and Error Analysis for the Physical Sciences},
Second Edition, (McGraw-Hill, Inc., New York, 1992).

\bibitem{Ieiri}
T. Ieiri, ``Measurement of Bunch Length Based on Beam Spectrum
at KEKB Rings,'' Proc. of EPAC 2000, Vienna, Austria, June 2000, p.~1735.

\end{thebibliography}
\end{document}